%% file: paper.tex
\documentclass[]{aa}
\usepackage{graphicx}
\usepackage[intlimits,sumlimits]{amsmath}
\usepackage{deluxetable}
\usepackage{times,epsfig} 
\usepackage{natbib}
\bibpunct{(}{)}{;}{a}{}{,} % to follow the A&A style
\usepackage{amsmath}
\usepackage[english]{babel}

\newcommand{\beqa}{\begin{eqnarray}} 
\newcommand{\eeqa}{\end{eqnarray}}
\newcommand{\Abst}[1]{\;#1}
\newcommand{\bsub}{\begin{subequations}}
\newcommand{\esub}{\end{subequations}}
\newcommand{\beal}{\begin{align}}
\newcommand{\ealn}{\end{align}}
\newcommand{\Nif}{$\rm ^{56}Ni$} 
\newcommand{\ksm}{${\rm km~s^{-1}~Mpc^{-1}}$}
\begin{document}

\title{Lower limits on the Hubble Constant from models of Type~Ia Supernovae}
\titlerunning{SNe~Ia and the Hubble Constant}

\author{M.~Stritzinger\inst{1} \and B.~Leibundgut\inst{2}}
\authorrunning{M. Stritzinger \and B. Leibundgut}

\offprints{\\M. Stritzinger, \email{stritzin@mpa-garching.mpg.de}}

\institute{Max-Planck-Institut f\"ur Astrophysik, Karl-Schwarzschild-Str. 1,
85741 Garching bei M\"unchen, Germany
\\e-mail: stritzin@mpa-garching.mpg.de
%-----------------
\and
%-----------------
European Southern Observatory, Karl-Schwarzschild-Str. 2, 85748
Garching bei M\"unchen, Germany
\\e-mail: bleibundgut@eso.org}

\date{Received; Accepted }

\abstract {By coupling observations of type~Ia supernovae with results
obtained from the best available numerical models we constrain the Hubble 
constant, independently of any external calibrators. We find an absolute 
lower limit of $H_{\circ}$ $>$ 50~\ksm. In addition, we construct a Hubble 
diagram with UVOIR light curves of 12 type~Ia supernovae located in the 
Hubble flow, and when adopting the most likely values (obtained from 1-D 
and 3-D deflagration simulations) of the amount of \Nif~produced 
in a {\it typical} event, we find values of $H_{\circ}$ $\geq$ 66$\pm$8 
and $\geq$ 78$\pm$9 \ksm, respectively. 
Our result may be difficult to reconcile with recent discussions in 
the literature as it seems that an Einstein-de Sitter universe 
requires $H_{\circ}$$\simeq$ 46~\ksm in order to fit the 
temperature power spectrum of the cosmic microwave background
and maintain the age constraints of 
the oldest stars. 
%The advantage of our method is that with very few 
%simple assumptions we bypass all the problems associated with the 
%extragalactic distance ladder and previous determinations of $H_{\circ}$.

%-----------------
}

\maketitle
\section{Introduction}

Due to their high intrinsic luminosity and apparent uniformity, 
type~Ia supernovae (hereafter 
referred to as SNe~Ia) have become an important distance indicator 
in modern cosmology. Today, they are utilized to determine the value 
of the Hubble constant ($H_{\circ}$) and measure the changes of the past cosmic 
expansion. For this reason a number of substantial observing campaigns 
have recently been conducted for SNe~Ia at nearby redshifts
\citep[see][ for a list]{leibundgut00}. As a 
result there is now a considerable number of events available with 
superb temporal and photometric coverage. However, there has been little 
effort made to use these high quality data sets to link observations with 
the physics of SNe~Ia in a systematic way. The purpose of this article 
is to combine results obtained from theoretical models 
with modern data in order to constrain the value of $H_{\circ}$.

Prior attempts to couple observations with explosion models of SNe~Ia in order 
to determine the value of $H_{\circ}$ include the pioneering
investigations of 
\citet{arnett85}, followed by \citet{branch92}, \citet{leibundgut92}, 
\citet{nugent95} and \citet{hoflich96}. 
These works gave promising results, constraining the Hubble
constant between 45~$\leq~H_{\circ}~\leq$~105~\ksm, and revealed 
that with few assumptions, SNe~Ia used in such a manner provide an 
attractive way to measure $H_{\circ}$, while circumventing many problems 
associated with the extragalactic distance ladder \citep[see][]{bohm97,livio97}.

Although several progenitor models for SNe~Ia are discussed, the common
view favors an accreting carbon oxygen (C-O) white dwarf in a binary
system, which undergoes thermonuclear incineration at or near the
Chandrasekhar mass \citep[for reviews
see][]{hillebrandt00,arnett96,woosley86}. The energy released from
burning to nuclear statistical equilibrium (NSE) at the density and
temperatures expected in these explosions completely disrupts the C-O
white dwarf, while the subsequent light curve is powered by the
Comptonization or ``thermalization'' of $\gamma$ rays produced from the
radioactive decay of $^{56}$Ni $\rightarrow$ $^{56}$Co $\rightarrow$
$^{56}$Fe. 
The single degenerate model became favored over the double
degenerate model after strong H$\alpha$ emission 
was observed in SN~2002ic \citep{hamuy03}. The H$\alpha$ emission 
in this SN~Ia is thought to come from circumstellar material 
originating from the
white dwarf's evolved companion star \citep{hamuy03,nomoto04}.

Here we use bolometric light curves of SNe~Ia as a means to link
observations with results obtained from models of an exploding C-O white
dwarf. Bolometric light curves constructed from observations provide a
simple and direct route to probe the complicated explosion physics and
radiation transport. As it is typically more straightforward to extract
the total flux (hence luminosity) of a SN~Ia from models rather than
the flux for individual filters, which require complicated multi-group
calculations \citep{leibundgut92,eastman97,hoflich97,leibundgut00}
bolometric light curves provide a sorely needed tool to connect
observations with models. In addition, observed bolometric light curves
are easily constructed from the integration of broad-band photometry, and
near maximum light they reflect the fraction of $\gamma$ rays thermalized.
Consequently $\sim$80\% or more
\citep{suntzeff96, suntzeff03} of the thermalized flux from the
$\gamma$ rays is emitted at optical and near-infrared wavelengths
(3000-10000~\AA), therefore what we manufacture from the observed
photometry has been coined a {\it UVOIR} bolometric light curve. 
It must be noted that by neglecting the small amount of flux 
outside the UVOIR wavelength regime we introduce a systematic 
underestimation on the calculated values of H$_{\circ}$. However, we 
address this systematic error when placing constraints on 
H$_{\circ}$ (see below). 
The summed UVOIR flux also offers the advantage that we do not need to apply
any K-corrections, which are necessary when just using individual filter
passbands of SNe~Ia located in the Hubble flow. Essentially, we can
reduce the problem to one of energy balances where the energy inputs
from the radioactive decays and the losses due to $\gamma$ ray escape
can be compared to the observed wavelength-integrated flux of the
SNe~Ia.
 
In the following we utilize a set of well observed SNe~Ia to demonstrate,
via two methods, (see below $\S$ \ref{method1} and $\S$ \ref{method2})
that --under the assumption that SNe~Ia are a product of the thermonuclear
disruption of a Chandrasekhar-mass C-O white dwarf-- it proves to be
rather difficult to obtain a value of $H_{\circ}$ $<$ 50~\ksm. Our
results, along with recent detection of the integrated Sachs-Wolfe (ISW)
effect \citep{boughn04a,boughn04b} observed in the Wilkinson Microwave
Anisotropy Probe (WMAP) data, \citep{bennett03,spergel03} cast doubts on
recent discussions in the literature
\citep[see][]{blanchard03,shanks04}, which suggest ``alternatives'' to
the concordance cosmological model. Spatially flat, matter-dominated
Einstein-de~Sitter models may produce a temperature power spectrum that
can fit cosmic microwave background (CMB) observations just as accurately
as the best concordance model, which sports a dark energy
component. However, Einstein-de~Sitter models require very low values of
the Hubble constant (e.g. $H_{\circ}$ $\simeq$ 46~\ksm) and are unable
to account for the observed ISW effect.

The structure of this paper is as follows: in $\S$~\ref{data} we briefly
discuss the basic details and information that we can extract from a
UVOIR/bolometric light curve. This is followed by a short discussion of
the data we have compiled and the method with which we construct our
UVOIR light curves. In $\S$~3 we discuss different models of SNe~Ia and
the \Nif~yields we adopt for a typical SN~Ia. We then present in
$\S$~\ref{method1} a method to derive $H_{\circ}$ through the
combination of observations and the theoretical \Nif~masses 
calculated in explosion models. Section \ref{method2} contains a
discussion of the classical way to derive $H_{\circ}$ through the
Hubble diagram of SNe~Ia. Contrary to previous methods we employ here
the bolometric flux. We conclude in $\S$~\ref{discussion}.

%---------------------
%\newpage
\section{Bolometric Lightcurves of SNe~Ia}
\label{data}
\subsection{UVOIR Bolometric Lightcurves}

Here we provide a basic description of a typical UVOIR light curve and
the physics that is thought to be driving its evolution and specific
characteristics at different epochs. During maximum light the dynamical
time (i.e. time since explosion) is approximately equal to the diffusion
time for photons trapped within the ejecta. This causes a reduction in
the opacity, which then allows a larger fraction of photons to escape
the expanding ejecta \citep{pinto01}. The majority of the observable
supernova flux is emitted in the optical. After maximum light an ever
increasing fraction of $\gamma$ rays escape freely and no longer deposit 
their energy in the ejecta. They are lost in the energy balance.
In addition, due to a lack of near-infrared 
($JHK$-passbands) data, we neglect a small contribution of flux 
(no more than $\sim$5\% near maximum light) in our constructed 
UVOIR light curves \citep{contardo01}. Right after maximum light as both the 
$\gamma-$ray deposition rate and the temperature, hence opacity, decrease, 
it is believed 
that there is a release of `old' photons, causing the observed luminosity 
to briefly overshoot the energy input from radioactive decay. 
After the release of stored energy the light curve declines in luminosity 
until between 20 and 40 days after maximum light, where an inflection point 
is seen in most events (see \citet{suntzeff96} and \citet{contardo00}). 
After $\sim$60 days the bolometric light curve begins to follow a nearly
linear decline of 0.026$\pm$0.002 mag/day \citep{contardo00} as the
energy input from radioactive decay decreases exponentially and less
energy is deposited by the $\gamma$ rays. 
At this time the infrared contribution rises to around 10\% 
\citep{contardo00}. About a year after maximum
the nebula no longer traps any $\gamma$ rays and the UVOIR light
curve is completely powered by positrons \citep{milne99}.

With UVOIR light curves and accurate distances we are able to obtain a
measure of the total luminosity and, through application of Arnett's
Rule, the quantity of $^{56}$Ni produced from burning to NSE
\citep{arnett82,arnett85}. Arnett's Rule simply states that during the
epoch of maximum light the luminosity of a SN~Ia is equal to the
instantaneous energy deposition rate from the radioactive decays within
the expanding ejecta \citep[see also][]{pinto00a,pinto00b}. This rule
has been utilized by \citet{suntzeff96}, \citet{vacca97},
\citet{contardo00}, \citet{strolger02}, and \citet{candia03} to
determine the amount of $^{56}$Ni produced in a number of SNe~Ia. These
efforts have revealed that the explosions of SNe~Ia do indeed produce
a range in the amount of $^{56}$Ni synthesized from $\sim$0.1
$M_{\sun}$ associated with the subluminous variety of SNe~Ia to
$\approx~$1 $M_{\sun}$ for the most luminous ones. 
We are now in the position to use UVOIR light curves of a
fair sample of well-observed SNe~Ia to probe the explosion mechanism.
Hopefully in the near future it will be possible to place constraints on 
models as they become more sophisticated. 
In a subsequent paper we will provide a detailed 
analysis of the bolometric light curves and derived \Nif~masses for a large 
number of well observed SNe~Ia.

\subsection{Observational data}
\label{obs}

As mentioned above, there are a number of past (and present) dedicated
monitoring programs located around the world that have assembled large 
collections of SNe~Ia data. Programs which we 
have used here include: the Cal\'{a}n/Tololo Survey \citep{hamuy95,hamuy96c}, 
the Center for Astrophysics \citep{riess99a,jha02}, and the Supernovae Optical 
Infrared Survey (SOIRS) \citep{hamuy01}. 

We selected only SNe~Ia located in the Hubble flow ($\geq$~3000 km s$^{-1}$) 
with excellent $(U)BVRI$-band observations and that
contain at least two pre-maximum observations in most photometric bands. Four 
of the SNe~Ia compiled here include $U$-band photometry, and for those
events without $U$-filter observations
we added a correction (see below). At this stage, no corrections were 
made to account for contributions by $UV$-flux blueward of the atmospheric 
cutoff and near-infrared $JHK$-band photometry. 

Table~1 lists all the SNe~Ia (and references) we have used
along with information pertaining to the amount of reddening that we
have adopted for each event. Values listed for Galactic reddening were
taken from the COBE dust maps of \citet{schlegel98}, while host galaxy
reddenings were procured from a variety of literature sources. To be as
consistent as possible we used reddenings given in \citet{phillips99}
for all SNe~Ia that coincided with our sample. For those events not
included in \citet{phillips99} we adopted values from the literature
giving preference to those calculated via the Phillips method. The
reddening for the host galaxy of SN~1999dq was taken from
\citet{riess04}. 

Finally, in Table~1 we list the two observables that are 
employed to constrain H$_{\circ}$. This includes
the host galaxy recession velocity and the UVOIR bolometric flux
at maximum light. Heliocentric 
velocities obtained from NED were converted to the CMB frame. 
As all SNe~Ia listed in Table~1 are located in the
Hubble flow, we assumed an error of 400 km s$^{-1}$ for all velocities, in 
order to account for (random) peculiar motions.
The uncertainties listed with the bolometric fluxes account for (1) a small 
measurement error, which is less than 5\% and (2) the uncertainties associated 
with estimates of host galaxy extinction.

\subsection{Construction of UVOIR Lightcurves}

We construct UVOIR light curves in the same manner previously adopted by 
\citet{vacca96, vacca97}, \citet{contardo00} and \citet{ contardo01}.
The reader is referred to these papers for a detailed discussion 
of this empirical fitting method and their previously attained results; 
here we briefly summarize the main points. 

We attempt to fit SNe~Ia photometry in a completely objective way. Data
for each filter is fitted with a ten parameter function. This function
consists of a Gaussian, corresponding to the peak phase on top of a
linear decline for the late time decay, an exponentially rising function
for the initial rise to maximum, and a second Gaussian for the secondary
maximum in the $VRI$ light curves. Fitting photometry in this manner is
advantageous because a continuous representation of the light curves is
produced without resorting to templates that may wash out subtleties of
each filtered light curve. The ten fitted parameters and several other
interesting quantities can be used to explore the finer details of
SNe~Ia light curves \citep[see][]{contardo01,stritzinger05}. 

To produce a UVOIR light curve we first fit the light curve of each
passband. Each magnitude is then converted to its corresponding flux at
the effective wavelength and a reddening correction is applied. The flux
for each filter at a given epoch is then integrated over wavelengths to
get the total flux. Note, corrections are employed to account for
overlaps and gaps between passbands. We also included a compensation in
a manner similar to \citet{contardo00} for those SNe~Ia that have no
$U$-band photometry. \citeauthor{contardo00} used a correction based on
SN~1994D \citep{richmond95,patat96,meikle96,smith00}, however, this
event had an unusually blue color at maximum and corrections based on it
tend to overestimate the fraction of flux associated with the $U$-band
photometry. We instead employed a correction derived from SN~1992A
\citep{suntzeff96}, which is the only well observed {\it normal} SN~Ia
with no host galaxy reddening. We estimate an additional 2$\%$ error is
incurred on each UVOIR light curve that has our $U$-band correction. 

%---------------------
\section{\Nif~Yields From Explosion Models}
\label{models}

A key ingredient for the methods presented below (see \S~\ref{method1}
and \S~\ref{method2}) is the amount of \Nif~produced in a {\it typical}
SN~Ia explosion. Both methods depend on the total energy radiated by the
supernova to establish its distance. 
\citet{contardo01} showed for a small sample of SNe~Ia that 
up to a factor of 10 difference in the yield of \Nif~can exist between
individual events. An absolute upper
limit for the amount of \Nif~synthesized is the Chandrasekhar mass 
($\sim$1.4 M$_{\sun}$), when the star becomes unstable and either
collapses or explodes.
However, due to the presence of intermediate mass elements (IMEs)
observed in spectra taken near maximum light, we know that the white
dwarf is not completely burned to \Nif. A lower limit is provided
by the subluminous events. Although only a few of these SNe~Ia have
been observed in detail (due to selection effects) three well 
observed events indicate $\sim$0.10 M$_{\sun}$ of \Nif~is synthesized
\citep{stritzinger05}.

To obtain a more quantitative value we turn our attention to recent
nucleosynthesis calculations performed at the Max-Planck-Institut f\"ur
Astrophysik (MPA) \citep{travaglio04}, which are based on 3-D Eulerian
hydrodynamical simulations \citep{reinecke02a,reinecke02b} of an
exploding white dwarf, that burns via a purely turbulent deflagration
flame.\footnote{Note that in 3-D deflagration simulations, once the 
initial conditions are set (i.e. T, $\rho$, and chemical composition) 
the only parameter that may be adjusted is the manner in
which the flame is ignited. Thus the amount of material burned is
determined by the adopted sub-grid model and the fluid motions on the
resolved scales \citep{reinecke02a}. Unlike 1-D simulations it is
impossible to fine tune the amount of material burned at a given 
density.} Based on their highest resolution 3-D simulation (i.e. model
b30\_3d\_768), which consisted of 30 ``burning'' bubbles and a grid size
of 768$^{3}$ for 1 octant of a sphere, \citeauthor{travaglio04} found 
the total yield of \Nif~to be 0.42~M$_{\sun}$. However, they found 
that as the number of ignition spots is increased, more explosion energy is
liberated, which may lead to a larger yield of \Nif. The
number of these ignition spots is strictly dependent on the grid
resolution of the simulation, which is limited by the computational
facilities available. One may therefore expect a larger yield
of \Nif~as the computational power and hence grid resolution is
increased. In addition, more recent calculations that employ ignition
conditions representing a foam-like structure, consisting of
overlapping and individual bubbles, indicate that it may be possible to
liberate a larger fraction of nuclear energy as one employs
different ignition conditions \citep{ropke04}.

We considered a range of results produced by other deflagration models
available in the literature, in particular the phenomenological
parametrized 1-D model - W7 \citep{nomoto84}. Recent nucleosynthesis
calculations show that W7 synthesizes 0.59 M$_{\sun}$ of \Nif~
\citep{iwamoto99}. It must be noted, however, that 1-D models compared
to 3-D calculations are expected to provide a less realistic
representation of the physical processes that occur during thermonuclear
combustions because they do not properly model the turbulent flame
physics. Additionally, multidimensional effects are neglected which do
have an important influence on the flame propagation. Nevertheless, W7
is a well established model that can fit the observed spectra rather well
\citep{harkness91,mazzali95,mazzali01} and has been used extensively over 
the last two decades to investigate SNe~Ia explosions.

Despite the success of deflagration models, they are currently unable to
account for the more luminous SNe~Ia, and predict appreciable
amounts of unburned carbon, oxygen, and IMEs leftover in the inner ashes
of the ejecta, which has not yet been conclusively observed. These
shortcomings were the motivation for the delayed detonation models
(DDM) \citep{khokhlov91,woosley90,woosley94, hoflich96}. In these models
the explosions starts as a deflagration flame until a transition 
occurs, causing the flame
to propagate supersonically thus the explosion becomes a detonation. 
\citet{hoflich95} provided a series of DDMs that range in \Nif~mass
between $\sim$0.34 and 0.67 M$_{\sun}$. His best fit model (M36) for
the well observed SN~1994D produces 0.60 M$_{\sun}$ of \Nif. The main
difference to the pure deflagration models is that DDMs contain an
additional free parameter that describes the local sound speed ahead of
the flame. This free parameter is not physically understood but is
essential to force the transition from deflagration to detonation. 

Throughout the following analysis we adopt results from the highest
resolution MPA simulation and the 1-D model W7. 
Although these two models are
not meant to represent the complete range in observed luminosity 
for the total population of SNe~Ia, they produce results that are 
illustrative of the majority of observed events. Both of these models 
are not perfect and, as results shown below indicate, the MPA 
model may not be representative of the more luminous events. 
We take the \Nif~ masses of these two models to be representative of a 
fair fraction of observed SNe~Ia. 

As previously noted in $\S$~1, just
after maximum light the observed luminosity is expected to be larger
than the radioactive luminosity, as the ejecta becomes optically thin and
allows the release of stored UVOIR photons. At this epoch
the photosphere rapidly recedes into the ejecta, revealing deeper layers
of the progenitor allowing more spectral lines to radiate.
\citet{branch92} (see his Table 1) conducted a survey of the best
numerical models at the time and found that the models which adequately
treat the time dependent nature of the opacity near maximum light
predict $\alpha$ (the ratio of energy radiated at the surface to the
instantaneous energy production by the radioactive decays) to be
slightly larger than unity. He concluded that $\alpha$$=$1.2$\pm$0.2
was the most applicable value and noted that the value of $\alpha$
appeared to be independent of the rise time. 

The parameter $\alpha$$=$1.2$\pm$0.2 is nothing more than a correction factor 
that is applied to the measured luminosity derived from the model \Nif~masses. 
In the following, we take this parameter into account in our discussion
of the values of H$_{\circ}$ determined from the models. 
The \Nif~masses of 0.42 and 0.59~M$_{\sun}$ correspond to an energy 
release after 19$\pm$3 days (the typical rise time of SNe~Ia \citep{contardo00})
of (8.40$\pm$1.26$)\times$10$^{42}$~erg~s$^{-1}$ and 
(1.18$\pm$0.18)$\times$10$^{43}$~erg~s$^{-1}$, respectively. If we combine 
the energy production with $\alpha$$=$1.2$\pm$0.2, the 
luminosity is increased to
 1.01$\pm$0.23$\times$10$^{43}$~erg~s$^{-1}$ 
and 1.42$\pm$0.32$\times$10$^{43}$~erg~s$^{-1}$. We note, however, that 
radiation transport 
calculations based on two MPA 3-D simulations (\citeauthor{blinnikov04}, 
private comm.) give the same value of $\alpha$ as calculated by Arnett's 
more simple analytical models (i.e. $\alpha$$=$1). 

Finally, we note that the value of $\alpha$ may vary for different
SNe~Ia depending on the amount and distribution of radioactive isotopes,
i.e. the opacity, in the ejecta. If $\alpha$$=$1 
occurs before bolometric maximum, one would expect a smaller amount of stored
photons. Therefore the luminosity would tend not to overshoot the
instantaneous energy deposition rate. However, if $\alpha$$=$1 occurs
{\it after} bolometric maximum, the light curve should in principle, be 
broader and flatter.

\section{H$_{\circ}$ from model \Nif~masses}
\label{method1}
In this section we derive a first analytic expression to constrain
$H_{\circ}$ directly from model \Nif~masses. 
This expression combines the UVOIR peak brightness of SNe~Ia with
explosion models via Arnett's Rule \citep{arnett82}. 
%---------------------
\subsection{Connecting $H_{\circ}$ and the Model Luminosities}
\label{eq1}

First, we develop an analytic equation which uses a simple argument that 
allows one to connect 
$H_{\circ}$ with the amount of \Nif~produced in a SN~Ia explosion. 
This relation relies on the fact that $H_{\circ}$ is defined as the
ratio of the local expansion velocity to the luminosity 
distance, which in turn is obtained from the inverse square law for the 
ratio of luminosity and the observed brightness. Therefore since the 
luminosity of a SN~Ia depends on the amount of \Nif, we can use the explosion 
models as our guide to the absolute luminosity. Combining this with 
both the measured brightness and recession velocity (or redshift) of any 
particular event, we can derive a value for $H_{\circ}$. The first 
expression to constrain $H_{\circ}$ therefore combines three elements: 
(1) Hubble's law of local cosmic expansion, (2) the distance 
luminosity relation, and (3) Arnett's Rule. 

We combine Hubble's law, which is defined by

\beal
\label{eq:hublaw}
H_{\circ} = \frac{cz}{D_{L}}
%\Abst{,}
\end{align}
with the distance luminosity relation given by 
\beal
\label{eq:dlr}
D_{L} = \left(\frac{L}{4 \pi F}\right)^{\frac{1}{2}}
\Abst{,}
\end{align}

\noindent where $F$ corresponds to the UVOIR flux obtained from
the bolometric light curves, $L$ is the luminosity of a
{\it fiducial} SN~Ia, and $D_{L}$ is the luminosity distance. We obtain
\beal
\label{eq:hub2}
H_{\circ} = cz \left(\frac{4 \pi F}{L}\right)^{\frac{1}{2}}
\Abst{.}
\end{align}

At maximum light the luminosity produced by the radioactive \Nif~ can be 
expressed as
\beal
\label{eq:lum}
L_{max} = \alpha E_{{\rm Ni}}(t_{R})
\Abst{,}
\end{align}
where $E_{{\rm Ni}}$ is the energy input from the decay of \Nif, evaluated
at the time of bolometric maximum (rise time $t_{R}$), and $\alpha$
accounts for any deviations from Arnett's Rule (where $\alpha$$=$1).
An expression for $E_{{\rm Ni}}$ can be found in \citet{nadyozhin94}

\begin{multline}\label{eq:E1}
E_{{\rm Ni}}(t_{R}) = \frac{\lambda_{{\rm Ni}}\lambda_{{\rm Co}}}{\lambda_{{\rm Ni}} - \lambda_{{ \rm Co}}}
\cdot \bigg\{ 
\left[Q_{{\rm Ni}} \left ( \frac {\lambda_{{\rm Ni}}}{\lambda_{{\rm Co}}} - 1 \right) - Q_{{\rm Co}}\right]
\bigg.
\\[2mm]
\bigg.
\cdot e^{- \lambda_{{\rm Ni}} t }+ Q_{{\rm Co}} e^{-\lambda_{{\rm Co}} t} 
\bigg\} M_{{\rm Ni}} = \epsilon(t_{R})M_{{\rm Ni}}
\Abst{,}
\end{multline}

\noindent where $\lambda_{{\rm Ni}}$ and $\lambda_{{\rm Co}}$ are e-folding 
decay times of 8.8 and 111.3 days for $^{56}$Ni and $^{56}$Co, respectively, 
and $Q_{{\rm Ni}}$ and $Q_{{\rm Co}}$
correspond to the mean energy released per decay of 1.75 and 3.73 MeV.
For 1 $M_{\sun}$ of \Nif, equation \eqref{eq:E1} turns into
\beal
\label{eq:E2}
E_{{\rm Ni}}(1~M_{\sun}) = 6.45 \times 10^{43}e^{-t_{R}/8.8} + 1.45 \times 10^{43}e^{-t_{R}/111.3}
\Abst{.}
\end{align}

\citet{riess99b} found, for a normal SN~Ia (e.g.~$\Delta m_{15}(B)$$=$1.1 mag) 
with a peak magnitude $M_{V}$~$=$~$-$19.45,
a rise time to $B$~maximum of $\sim$19.5 days. \citet{contardo00} found 
the bolometric rise time to be within one day of the $B$-band for nearly all 
SNe~Ia in their sample. Throughout this work we assume a bolometric rise time of 
19$\pm$3 days. The adopted uncertainty should be adequate to account for 
intrinsic differences between the rise times of different SNe~Ia.
Using this rise time and assuming $\alpha$$=$1.0 (Arnett's Rule) we can combine
equations \eqref{eq:lum} and \eqref{eq:E2} and obtain the simple 
relation that gives for 1 M$_{\sun}$ of \Nif~ a total luminosity at maximum 
light of 
\beal
\label{eq:hub3.0}
{\rm L_{max}}(1~M_{\sun})~=~(2.0\pm0.3)~\times~10^{43}~{\rm erg~s^{-1}~M^{-1}_{\sun}}
\Abst{,}
\end{align}
where the error corresponds to the 3 day uncertainty in the adopted 
bolometric rise time.

Substituting equation \eqref{eq:lum} into equation \eqref{eq:hub2} we can relate
the luminosity to the mass of \Nif~via $E_{{\rm Ni}}$
(equation~\ref{eq:E1}), and then if we include the factors
that directly equate the luminosity with the \Nif~mass, 
we obtain our final relation to calculate the Hubble constant 
\\
\beal 
\label{eq:hubequal}
H_{\circ} = cz \left(\frac{4 \pi F_{max}^{bol}}{L_{max}}\right)^{\frac{1}{2}}~
=~~ cz \left(\frac{4 \pi F_{max}^{bol}}{\alpha E_{\rm Ni}(t_{R})}\right)^{\frac{1}{2}} = cz \left(\frac{4 \pi F_{max}^{bol}}{\alpha \epsilon(t_{R}) M_{\rm Ni}}\right)^{\frac{1}{2}} 
\Abst{.}
\end{align}
\\

With equation \eqref{eq:hubequal} only two observables -- the bolometric
flux ($F_{\rm max}^{bol}$) at maximum and the redshift ($z$) -- are required
to determine the value of $H_{\circ}$. 

Uncertainties come from the rise time, which determines the
the peak luminosity, the uncertainty in $\alpha$, which depends on the
radiation escape from the explosion, and of course the amount of 
\Nif~synthesized 
in the explosion. Finally we note that our \Nif~mass is 
``error free'' in the sense that the adopted value(s) for this parameter 
(hence fiducial luminosities) are completely model dependent.

We now have an analytic form for H$_{\circ}$ which is directly
connected to the \Nif~produced in the explosion. The other parameters
have to do with the radiation transport: the ratio of energy release to
energy input and the time between explosion and maximum luminosity. 

\subsection{Results}
\label{method1results}

In Fig.~1 we present results obtained using equation
\eqref{eq:hubequal} for all SNe~Ia listed in Table~1. For
every supernova we show the derived H$_{\circ}$, assuming that its
observed brightness would correspond to a given nickel mass
(in steps of 0.1~M$_{\sun}$). The inverse
square-root dependence of H$_{\circ}$ on the nickel mass is clearly
visible. The `1-$\sigma$' error bars that accompany each point account
for a recession velocity error of 400 km s$^{-1}$, an 
error associated with the reddening correction, 
a measurement error of the flux ($\leq$~5\%), a 3 day error for
the assumed bolometric rise time, and a 2\% error for those events that
have a $U$-band correction. Note that the most dominant error is 
the uncertainty associated with the redshift due to peculiar velocities.

It is evident from this figure that for a given mass of \Nif~there 
exists a range of possible values of $H_{\circ}$. 
This is what we expect owing to the fact that there are known intrinsic 
differences between SNe~Ia. Hence, if we (erroneously) assume a single 
\Nif~mass for all observed SNe~Ia, we obtain a range of H$_{\circ}$ values 
as in Fig.~1. 

Two SNe~Ia (SN~1992bo and SN~1993H) are clearly situated below the rest
of the objects and are both known to be red events with
$\Delta$$m_{15}(B)=$1.69 \citep{hamuy96c}. Both also show evidence of weak
Ti features in their spectra (\citeauthor{phillips04}, private comm.).
Because the models we have adopted in this work were designed for 
normal SNe~Ia and these two events are subluminous in nature, we exclude
them in the following discussion; however, they are included in the Hubble 
diagram (see \S~\ref{method2}).
\footnote{On a uniform distance scale SN~1992bo and SN~1993H are 
$\sim$40\% less luminous than the other objects and hence produce
correspondingly less \Nif~\citep{stritzinger05}.}
Given the prediction of an explosion model, we can now read off the
preferred value of the Hubble constant. Na\"ively, one would like to
take the mean of the distribution of the supernovae for a given \Nif~mass 
and derive a Hubble constant. However, the natural scatter of 
\Nif~\citep{contardo00,bowers97,cappellaro97} prevents us from doing this
because only one value is correct for a given event.
The best we can do now is to derive a lower limit for H$_{\circ}$ by
associating the nickel masses with the faintest supernovae and hence
obtain an underestimate of the Hubble constant. 

By choosing a \Nif~mass of 1~M$_{\sun}$ and associating it with 
faint SNe~Ia we clearly reach a lower limit for $H_{\circ}$. Note
that the highest \Nif~mass derived from a SN~Ia with a Cepheid distance
is 0.84~M$_{\sun}$ \citep{strolger02}.
The solid horizontal line in Fig.~1 indicates that 
evidently more than 1~M$_{\sun}$ of \Nif~must be produced in a normal 
SNe~Ia explosion in order to obtain $H_{\circ}$~$<$~50~\ksm. 
Note that the two dotted vertical lines indicate our adopted 
\Nif~masses with $\alpha$$=$1.0. 
For both cases an H$_{\circ}$ of more than 50~km~s$^{-1}$~Mpc$^{-1}$ is
favored. Only the two faint SNe~Ia fall well below this value.

To obtain an absolute lower limit on $H_{\circ}$, we present in Fig.~2
the least luminous normal event in our sample --
SN~1999aa -- with 1-$\sigma$ and 3-$\sigma$ confidence levels. This SN~Ia
produces values of $H_{\circ}$ that are 9\% below the mean determined
from all 10 normal events. For our adopted \Nif~masses with $\alpha$$=$1
we obtain a lower limit on $H_{\circ}$ with SN~1999aa (3-$\sigma$ away
from the calculated values) to be $\geq$  40~\ksm. 
The dashed line in Fig.~2 illustrates the effect of 
increasing $\alpha$ by 20\%. We see that this gives us an additional 
systematic uncertainty that would decrease $H_{\circ}$ by 9\%. 
The change incurred on the Hubble constant lies directly on top of the 
quoted lower 1$\sigma$ confidence level. Finally, we note 
from equation~\eqref{eq:hubequal} that by neglecting $\sim$10\% of the flux 
emitted outside of the optical, we systematically underestimate H$_{\circ}$ 
by 5\%.

We find that with white dwarfs as progenitors of SNe~Ia it is very
difficult to obtain a value of $H_{\circ}$~$<$~50~\ksm. With
1~M$_{\sun}$ of \Nif, one could expect $H_{\circ}$ $\sim$60~\ksm.
Observations and models for the most luminous events indicate that no
more than 1~M$_{\sun}$ of \Nif~is produced. With our adopted \Nif~masses
(0.42 M$_{\sun}$ $<$ \Nif~$<$ 0.59 M$_{\sun}$) we find 
from the 1-$\sigma$ error bars in Fig.~2
the Hubble constant to be constrain between 70$\pm$6 $\leq$ H$_{\circ}$
$\leq$ 83$\pm$7~\ksm. 

The problem can, of course, also be inverted to derive a possible
range of \Nif~mass given a value of H$_{\circ}$. This will be interesting to
constrain the \Nif~mass for models, should H$_{\circ}$ be known to high 
accuracy. For $H_{\circ}$~$\approx~$70~\ksm we find from Fig.~1
a range in the amount of \Nif~produced in a SNe~Ia explosion to be 
0.5~M$_{\sun}<$ \Nif~$<1.0$ M$_{\sun}$. 

\section{H$_{\circ}$ through the Hubble diagram of SNe~Ia}
\label{method2}
With this method we determine $H_{\circ}$ from the Hubble diagram in a
manner similar to what has been previously presented by
\citet{tammann90} and \citet{leibundgut92} \citep[see also][ for similar
applications]{sandage93,hamuy96c,phillips99,parodi00}. 
We note that this method is similar to the previous method, 
however, here $H_{\circ}$ is calculated in a more traditional manner. 
An analytic expression to constrain $H_{\circ}$ from our Hubble diagram is trivial
to derive from the distance luminosity relation. Solving equation
\eqref{eq:dlr} for $F_{max}^{bol}$ and then taking the logarithm of both
sides, we obtain
\beal
\label{eq:hub4}
\log(F_{max}^{bol}) = \log\left(\frac{L_{max}}{4 \pi D_{L}^{2}}\right)
\Abst{.}
\end{align}
Substituting $cz/H_{\circ}$ for $D_{L}$ and rewriting the right hand side
of equation~\eqref{eq:hub4} we obtain

\beal
\label{eq:hub5}
\log(F_{max}^{bol}) = -2\log(cz) + \log(L_{max}) - \log(4 \pi) + 2\log(H_{\circ})
\Abst{.}
\end{align}
There is a linear relation between $\log(F_{max}^{bol})$ and $\log(cz)$ as can be
seen in Fig.~3.
A linear regression to the data yields a slope of $2.01\pm0.25$, which
is fully consistent with the expected slope of 2 for a linear local
expansion derived in equation \eqref{eq:hub5}. From Fig.~3
it is also obvious that the two faint objects are outliers. 
If they are ignored, the fit sharpens
up to a slope of $1.97\pm0.10$. With a fixed slope to this linear
expansion value of $2$ we derive the y-intercept, which corresponds to 
\beal
\label{eq:hub6}
b = -\log(L_{max}) + \log(4\pi) - 2\log(H_{\circ})
\Abst{.}
\end{align}
Solving equation~\eqref{eq:hub6} for $H_{\circ}$, we arrive at our final expression
for the Hubble constant
\beal
\label{eq:hub7}
H_{\circ}= \left(\frac{4 \pi}{10^{b} L_{max}}\right)^{\frac{1}{2}}
\Abst{.}
\end{align}

\noindent The Hubble constant is now simply calculated by
plugging in the y-intercept, b, derived from the linear regression of the
Hubble diagram and a {\it fiducial} luminosity defined by our adopted models. 

\subsection{Results}
\label{method2results}

In Fig.~3 we present our Hubble diagram containing all 
SNe~Ia listed in Table~1. Error bars for all events account
for both a redshift uncertainty of 400 km s$^{-1}$ and the uncertainties 
associated with host galaxy reddening.
A weighted least-squares fit to the Hubble diagram (for all 12 SNe~Ia), with 
a fixed slope of $2$, yields $b$~$=$3.292$\pm$0.047. 
Plugging this into equation~\eqref{eq:hub7} 
along with our adopted \Nif~masses of 0.42~M$_{\sun}$ and 0.59~M$_{\sun}$
we find H$_{\circ}$ to be $\geq$ 85$\pm$7 and $\geq$ 
72$\pm$6~\ksm (1-$\sigma$ error) respectively. Accounting for $\alpha$$=$1.2$\pm$0.2 
we obtain lower values of the Hubble constant to be H$_{\circ}$ 
$\geq$ 78$\pm$9 and $\geq$ 66$\pm$8~\ksm, respectively. 

As an upper limit in the amount of \Nif~synthesized in the most luminous SNe~Ia 
explosion is expected to be $\approx~$1~M$_{\sun}$, we can use the
corresponding luminosity as a guide to obtain a lower limit
on the Hubble constant through the Hubble diagram. Thus, for 1~M$_{\sun}$ of 
\Nif~and $\alpha$$=$1 we obtain a value of H$_{\circ}$ $\geq$ 55$\pm$5 
\ksm. As in the previous method, we are underestimating 
H$_{\circ}$ by $\sim$5\%, due to flux outside the optical window. 
\section{Discussion}
\label{discussion}

Under the main assumptions: (1) that the progenitors of SNe~Ia are C-O
white dwarfs that explode at or near the Chandrasekhar mass, and (2) that 
the amount of \Nif~synthesized to first order determines the peak
luminosity, we are able to use results obtained from state-of-the-art
numerical simulations of explosion models to uniquely define the 
bolometric peak luminosity, and in concert with photometric observations 
constrain H$_{\circ}$. The
attractiveness of this approach stems from the ability to bypass
assumptions that are typically made when one attempts to determine
H$_{\circ}$, i.e. the extragalactic distance ladder and its accumulation
of error from rung to rung. We stress 
that our fitting method does not add any corrections to the data. In
other words we {\it do not} normalize the flux to any decline rate
relation (e.g. $\Delta$$m_{15}$ \citep{phillips99}, MLCS
\citep{riess96} or stretch \citep{perlmutter97}).

Uncertainties that marginally affect (in decreasing order of importance)
our results include: the abundances of peak Fe group metals -- stable
and radioactive -- produced in the explosion models, the redshift peculiar 
velocities of each 
SN~Ia, the total absorption, the assumed rise time to bolometric maximum, 
the exact nature of $\alpha$, which may slightly vary from SN to SN depending 
on the exact nature of the opacity and ionization structure of the expanding 
ejecta, and the amount of flux that we neglect outside of the optical window. 

It is still unclear what parameters effect the amount of \Nif~produced
in a SN~Ia explosion. Obvious candidates are the initial conditions
prior to explosion. These include the metallicity of the C-O white dwarf,
the central density and the ignition mechanism. If there is a
considerable fraction of alpha elements such as $^{22}$Ne within the
progenitor one would expect more stable isotopes such as $^{58}$Ni and
$^{54,56}$Fe to be produced from burning to NSE, thus reducing the \Nif~yield 
\citep[e.g.][]{brachwitz00}. A higher central density on the other 
hand would lead to a more robust explosion and hence an increased amount 
of \Nif. As discussed earlier the explosion mechanism is still uncertain 
and different deflagration and detonation scenarios produce different 
amounts of peak Fe group elements. Nevertheless with a larger \Nif~mass 
we obtain smaller values of H$_{\circ}$. Errors attributed to the adopted 
recession velocity and reddenings produce scatter in our Hubble diagram but 
have very little effect on our calculations of H$_{\circ}$ via
equations~\eqref{eq:hubequal} or \eqref{eq:hub7}. The $\pm$3 day 
departure from our adopted rise has no more than a 10\% effect 
on our calculations. A change in $\alpha$ by $\pm$20\% can affect 
H$_{\circ}$ by 9\%. Finally, we reiterate from 
equation~\eqref{eq:hubequal} 
that by neglecting $\sim$10\% of the flux emitted outside of the optical, 
we are underestimating H$_{\circ}$ by at least 5\% or correspondingly
more if more flux is unaccounted in our method. 

In $\S$ \ref{method2results} we determined a rather high value (85~\ksm)
for the Hubble constant when using results from the MPA model. This 
indicates that the amount of \Nif~produced in these 3-D deflagration 
simulations currently are on the low side. And indeed 
a large sample of SNe~Ia show that the average distribution of \Nif~mass is 
slightly higher ($\sim$~0.6~M$_{\sun}$)
\citep{stritzinger05}. This suggests that their models may need more fine tuning
in order to produce a larger amount of \Nif, and hence match observations more
accurately.
 
There have been many attempts since \citet{kowal68} presented his Hubble
diagram to exploit SNe~Ia to determine H$_{\circ}$. We refer the reader
to \citet{branch98} for a detailed review of previous works 
that attempt to calculate H$_{\circ}$ based on
SNe~Ia. He concluded, from methods based on physical considerations
similar to the methods presented in this work and methods which utilize
SNe~Ia that have been independently calibrated by Cepheids, a range in
the Hubble constant of 54~$\leq$~H$_{\circ}$~$\leq$~67 \ksm, with a
``consensus" on H$_{\circ}$~$=$~60$\pm$10~\ksm. More recent
investigations of \citet{suntzeff99} and \citet{jha02} give values of
H$_{\circ}$~$=$~64~\ksm. Finally, 
\citep{freedman01,spergel03,altavilla04} have all measured slightly
larger values of H$_{\circ}$~$\approx~$70~\ksm with 10\% accuracy. 

Another method independent of the extragalactic distance ladder which
combines X-ray imaging of galactic clusters with the Sunyaev-Zel'dovich
effect (SZE) has been recently used to place limits on H$_{\circ}$
\citep{myers97,mason01, jones03,reese02}. These works have provided the
detailed study of 41 clusters giving distances which yield an averaged
value of H$_{\circ}$~$\approx~
$61$\pm$3~\ksm \citep[for a review
see][]{reese03}. 

We find from both methods presented here that the Hubble constant must
be $>$~50~\ksm in order to be compatible with current supernova models.
In addition, we stress that this lower limit is based on the assumption
that 1~M$_{\sun}$ is an upper limit on the amount of \Nif~produced in a 
SN~Ia explosion, and not from our adopted models.
This result, along with other methods to measure H$_{\circ}$ using SNe~Ia
calibrated with Cepheids, SZE/X-ray distances and evidence of the ISW
effect, strongly suggest that we do not live in a matter dominated
universe without some form of cosmological constant or similar agent.

\begin{acknowledgements}
We thank the anonymous referee for many helpful comments that 
significantly improved the presentation of this paper.
M.S. acknowledges the International Max-Planck Research School
on Astrophysics for a graduate fellowship. M.S. is grateful for helpful 
conversations with Sergei Blinnikov, Wolfgang Hillebrandt, Gert H\"utsi, 
Kevin Krisciunas, Paolo Mazzali, Friedrich R\"opke, and Stefanie Walch. 
Special thanks to Nick Suntzeff for his hospitality, strong mentorship, 
and access to SN~1992A data. This research has made use of the NASA/IPAC 
Extragalactic Database (NED), which is operated by the Jet Propulsion 
Laboratory, California Institute of Technology, under contract with the 
National Aeronautics and Space Administration.

\end{acknowledgements}

\clearpage
\input{paper.tab1.tex}

\clearpage
\begin{figure}
\resizebox{\hsize}{!}{\includegraphics{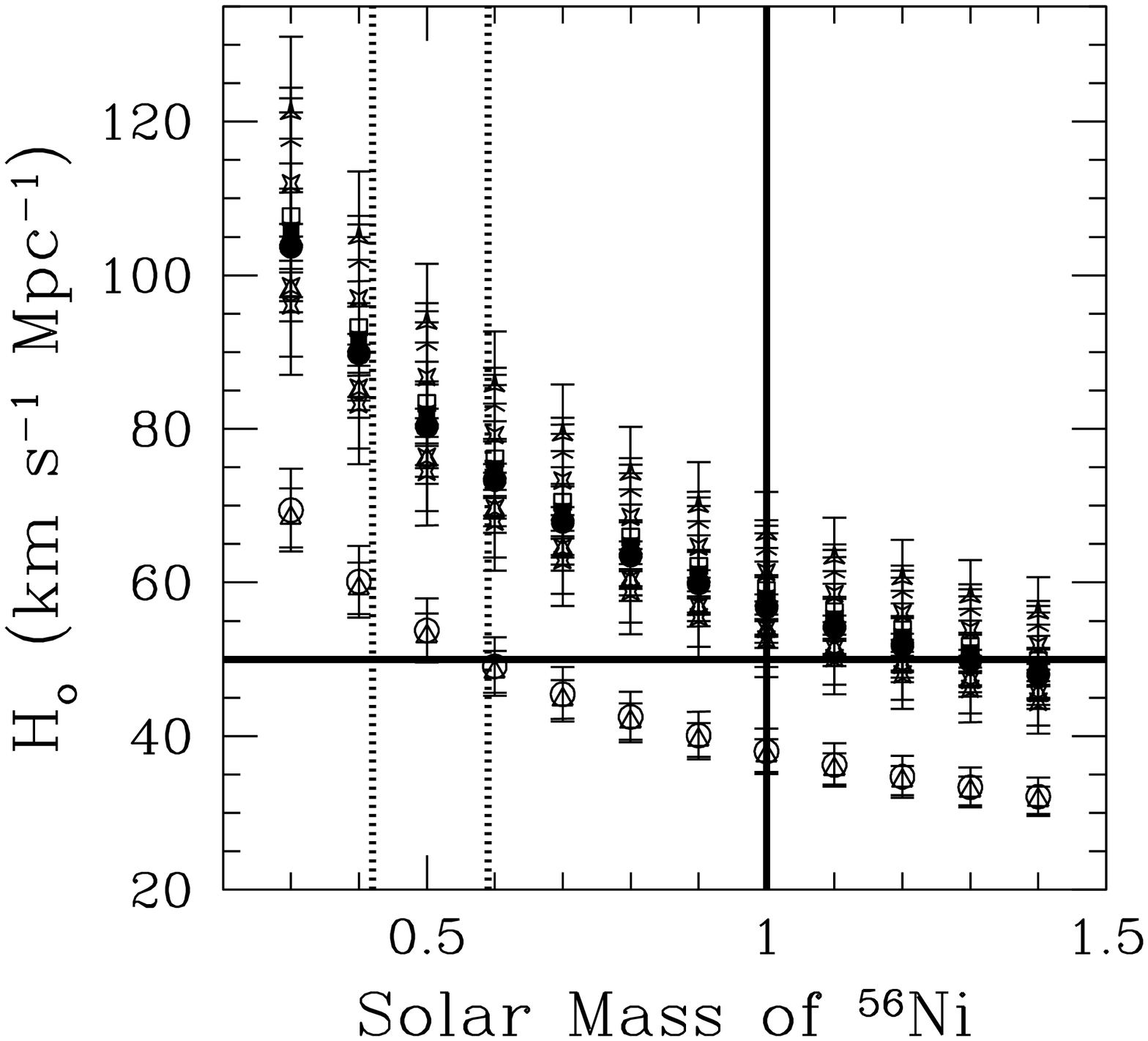}}
\caption{Hubble constant as a function of \Nif~mass for 12 SNe~Ia in the 
Hubble flow for $\alpha$$=$1. Error bars represent all uncertainties discussed 
in section \ref{eq1}.
The dotted lines indicate the two \Nif~masses from our adopted models.
The solid vertical line indicates an upper limit on the production of 
\Nif~expected
from the thermonuclear explosion of a C-O white dwarf, and the solid horizontal 
line corresponds to the lower limit we derive for H$_{\circ}$.}
\label{hubble.fig1}
\end{figure}

\begin{figure}
\resizebox{\hsize}{!}{\includegraphics{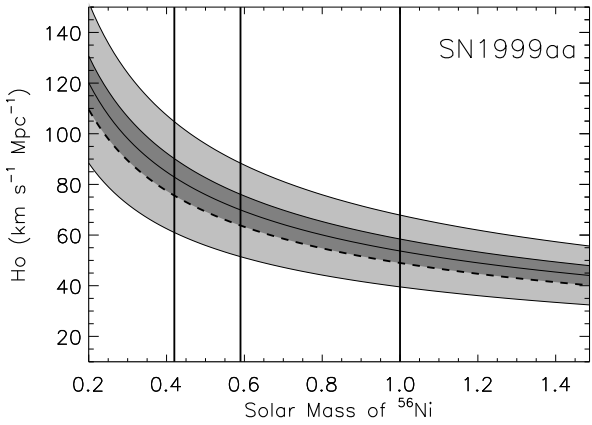}}
\caption{To highlight the absolute lower limit we derive on H$_{\circ}$ we plot results
obtained with SN~1999aa
(the least luminous of our 10 normal SNe~Ia) with 1-$\sigma$ (dark shading) and 3-$\sigma$ 
(light shading) confidence levels. Vertical solid lines indicate both the adopted \Nif~masses and 
the 1 M$_{\sun}$ upper limit expected from an exploding C-O white dwarf. 
The dashed line illustrates the effect if $\alpha$ is increased by 20\% (i.e. $\alpha$$=$1.2).}
\label{hubble.fig2}
\end{figure}

\begin{figure}
\resizebox{\hsize}{!}{\includegraphics{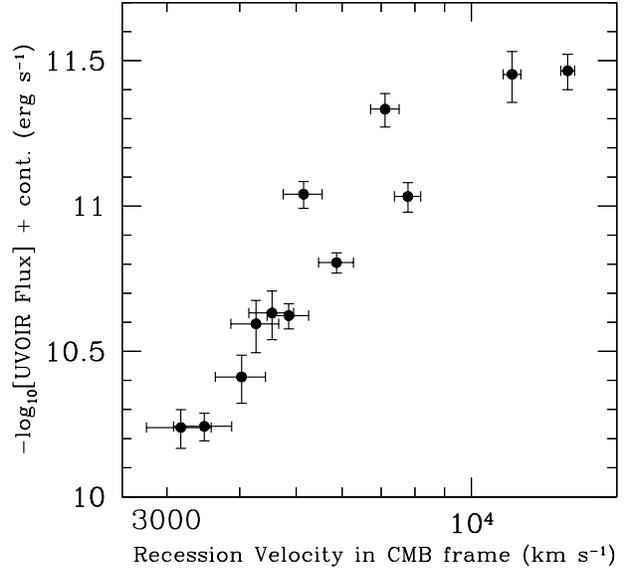}}
\caption{Hubble diagram of 12 SNe~Ia in the Hubble flow.
The negative logarithm of the UVOIR flux at maximum light is plotted
against the logarithm of the recession velocity in the CMB frame.
Error bars account for a peculiar velocity of 400 km s$^{-1}$ and 
uncertainties associated with host galaxy reddening.}
\label{hubble.fig3}
\end{figure}

\end{document}

%% file: paper.tab1.tex
\begin{deluxetable}{l l l c c c c}
\tablecolumns{7}
\tablenum{1}
\tablewidth{0pc}
\tablecaption{Well-observed SNe~Ia in the Hubble Flow}
\label{data.tab}
\tablehead{
\colhead{SN} &
\colhead{Filters} &
\colhead{Ref.} &
\colhead{E(B-V)$_{gal}^{a}$} &
\colhead{E(B-V)$_{host}$} &
\colhead{v$_{{\rm CMB}}$ $^{b}$} &
\colhead{F$_{max}^{bol}$} \\
\colhead{} &
\colhead{} &
\colhead{} &
\colhead{} &
\colhead{} &
\colhead{(km s$^{-1}$)} &
\colhead{(erg s$^{-1}$ cm$^{-2}$)}
} 
\startdata
SN1992bc & $BVRI$  & 1        & 0.022 & 0.000 & 5870  & (1.565$\pm$0.124)$\times 10^{-11}$ \\ 
SN1992bo & $BVRI$  & 1        & 0.027 & 0.000 & 5151  & (9.106$\pm$0.967)$\times 10^{-12}$ \\ 
SN1993H  & $BVRI$  & 1        & 0.060 & 0.050 & 7112  & (4.640$\pm$0.613)$\times 10^{-12}$ \\
SN1995E  & $BVRI$  & 2        & 0.027 & 0.740 & 3478  & (5.726$\pm$0.622)$\times 10^{-11}$ \\
SN1995ac & $BVRI$  & 2        & 0.042 & 0.080 & 14651 & (3.425$\pm$0.477)$\times 10^{-12}$ \\ 
SN1995bd & $BVRI$  & 2        & 0.495 & 0.150 & 4266  & (2.542$\pm$0.521)$\times 10^{-11}$ \\
SN1996bo & $BVRI$  & 2        & 0.078 & 0.280 & 4857  & (2.382$\pm$0.237)$\times 10^{-11}$ \\ 
SN1999aa & $UBVRI$ & 3, 4, 5  & 0.040 & 0.000 & 4546  & (2.333$\pm$0.446)$\times 10^{-11}$ \\
SN1999aw & $BVRI$  & 5, 6     & 0.032 & 0.000 & 11754 & (3.525$\pm$0.700)$\times 10^{-12}$ \\
SN1999dq & $UBVRI$ & 3        & 0.024 & 0.139 & 4029  & (3.871$\pm$0.730)$\times 10^{-11}$ \\
SN1999ee & $UBVRI$ & 7        & 0.020 & 0.280 & 3169  & (5.781$\pm$0.881)$\times 10^{-11}$ \\
SN1999gp & $UBVRI$ & 3, 8     & 0.056 & 0.070 & 7783  & (9.270$\pm$1.082)$\times 10^{-12}$ \\

\enddata
\tablenotetext{a} {Taken from \citet{schlegel98} dust maps.}
\tablenotetext{b} {Heliocentric velocities from NED transformed to the cosmic microwave background frame. To account for peculiar velocities we assume throughout this work an error of 400 km s$^{-1}$ for all CMB distances.}

\tablerefs{
(1)~- \citealt{hamuy96c}, (2)~- \citealt{riess99a}, (3)~- \citealt{jha02}, (4)~- \citealt{krisciunas00}, (5)~- \citealt{regnault00}, (6)~- \citealt{strolger02}, (7)~- \citealt{stritzinger02}, (8)~- \citealt{krisciunas01}}

\end{deluxetable}